\documentclass{article}

\usepackage{microtype}
\usepackage{graphicx}
\usepackage{float}
\usepackage{subcaption}
\usepackage{booktabs}
\usepackage{tabularx}
\usepackage{hyperref}
\usepackage{listings}
\usepackage[most]{tcolorbox}
\tcbuselibrary{listings,breakable}

\usepackage{amsmath}
\usepackage{amssymb}
\usepackage{mathtools}
\usepackage{amsthm}
\usepackage{algorithm}
\usepackage{algpseudocode}
\usepackage{fvextra} 

\makeatletter
\@ifundefined{crefname}{}{%
  \crefname{blocks}{Box}{Boxes}%
  \Crefname{blocks}{Box}{Boxes}%
  \crefname{interfaces}{Interface}{Interfaces}%
  \Crefname{interfaces}{Interface}{Interfaces}%
}
\makeatother

\usepackage[preprint]{icml2026}   

\usepackage[capitalize,noabbrev]{cleveref}
\icmltitlerunning{Tool-Augmented Agents for Lean Formalization}
\begin{document}
\twocolumn[
\icmltitle{Understanding Tool-Augmented Agents for Lean Formalization: A Factorial Analysis}


\begin{icmlauthorlist}
    \icmlauthor{Ke Zhang}{ucr}
    \icmlauthor{Patricio Gallardo}{ucr}
    \icmlauthor{Maziar Raissi}{ucr}
    \icmlauthor{Sudhir Murthy}{ucr}
\end{icmlauthorlist}
\icmlaffiliation{ucr}{University of California, Riverside}
\icmlcorrespondingauthor{Maziar Raissi}{maziar.raissi@ucr.edu}

\icmlkeywords{Tool-Augmented Agents, Lean 4, Formalization, Proof Assistants}
\vskip 0.3in
]
\printAffiliationsAndNotice{}

\begin{abstract}
Automatic translation of natural language mathematics into faithful Lean 4 code is hindered by the fundamental dissonance between informal set-theoretic intuition and strict formal type theory. This gap often causes LLMs to hallucinate non-existent library definitions, resulting in code that fails to compile or lacks semantic fidelity. In this work, we investigate the effectiveness of tool-augmented agents for this task through a systematic factorial analysis of three distinct tool categories: Fine-tuned Model Querying (accessing expert drafts), Knowledge Search (retrieving symbol definitions), and Compiler Feedback (verifying code via a Lean REPL). We first benchmark the agent against one-shot baselines, demonstrating large gains in both compilation success and semantic equivalence. We then use the factorial decomposition to quantify the impact of each category, isolating the marginal contribution of each tool type to overall performance.
\end{abstract}

\section{Introduction} \label{sec:intro}

Large language models (LLMs) have demonstrated remarkable capability in producing complex mathematical arguments~\cite{castelvecchi2025ai}, yet their probabilistic nature fundamentally conflicts with the absolute certainty required by formal mathematics. Proof assistants such as Lean 4~\cite{deMouraUllrich2021lean4} resolve this conflict by mechanically verifying logical correctness, but they introduce a significant barrier: the "formalization bottleneck." Most modern undergraduate and graduate-level theorems exist only in natural language and have not yet been translated into the strict syntax required by proof assistants~\cite{LeanToDo}.

Current attempts to automate translation into Lean 4 rely primarily on one-shot prompting or fine-tuning on static datasets \cite{gaoherald, wang2025kimina}. Although these methods can generate plausible code, they face several fundamental obstacles. Training data for graduate-level mathematics is sparse, causing models to hallucinate nonexistent theorems and symbols. Even when the mathematics is correct, a deep mismatch remains between the set-theoretic language of textbooks and Lean’s dependent type theory, making direct translation unreliable. Moreover, syntactic validity does not ensure semantic fidelity: Lean code may compile yet express a vacuous or incorrect statement. These issues are further compounded by the fact that Mathlib 4 is a fast-evolving, community-maintained library, so models trained on static snapshots frequently generate code that targets deprecated or nonexistent definitions in the current environment.

To address these limitations, we introduce an agentic framework that replaces static generation with iterative refinement. A generalist LLM orchestrator interacts with three tool categories—symbol retrieval, expert drafting, and Lean REPL feedback—forming a closed loop of draft, verify, and repair that corrects both syntactic and semantic errors. Rather than treating this toolset as a black box, we apply a full factorial design on a new benchmark of 400 graduate-level theorems to quantify the causal contribution and interaction of each component. This allows us to identify which tools create new capability regimes and which merely stabilize or accelerate convergence toward faithful formalizations.

Finally, to support reproducibility, we will open-source our entire research framework, including the full benchmark, agent implementation, evaluation scripts, and agent tool calling logs upon acceptance.

\section{Problem Setting}
\label{sec:problem}

We address the task of \textbf{formalization}: automatically translating natural-language mathematical statements into valid Lean 4 theorem declarations. This task serves as the critical first step in verifying mathematics, bridging the gap between informal human intent and machine-checkable syntax.

\subsection{Task Definition}
Formally, let $\mathcal{X}$ be the space of informal mathematical statements and $\mathcal{Y}$ be the space of valid Lean 4 source code. Given an input $x \in \mathcal{X}$, the system must generate a formal statement $y \in \mathcal{Y}$ that satisfies three criteria: it must possess \textbf{syntactic validity} (compile in Lean 4 with Mathlib), maintain \textbf{statement-centricity} (omitting proofs via \texttt{:= by sorry}), and ensure \textbf{semantic faithfulness} (logical equivalence to $x$).

\subsection{Dataset Construction}
To benchmark this task, we curated a dataset of 400 graduate-level theorems derived from open-source LaTeX lecture notes and textbooks. We selected sources that provide natural language statements without accompanying formal code, ensuring the task represents a genuine translation effort rather than retrieval. 

As detailed in Table~\ref{tab:dataset_sources}, we balanced the dataset across
four major mathematical domains—Real Analysis, Complex Analysis, Topology, and
Algebra—selecting 100 diverse examples from each to ensure broad coverage of
mathematical terminology and structures.

\begin{table}[t]
\centering
\small 
\caption{Benchmark Dataset Composition.}
\label{tab:dataset_sources}
\begin{tabularx}{\columnwidth}{@{} l >{\raggedright\arraybackslash}X c @{}}
\toprule
\textbf{Domain} & \textbf{Source Material} & \textbf{N} \\ \midrule
Real Analysis    & \textit{Basic Analysis} (Lebl)~\cite{lebl_basic_analysis_github} & 100 \\ \addlinespace
Complex Analysis & \textit{Cultivating Complex Analysis} (Lebl)~\cite{lebl_ca_analysis_github} & 100 \\ \addlinespace
Topology         & \textit{Notes on Topology} (McKay)~\cite{mckay_topology_notes_github} & 100 \\ \addlinespace
Algebra          & \textit{Abstract Algebra} (Doty)~\cite{doty_abstract_algebra_notes_github} & 100 \\ \midrule
\textbf{Total}   & & \textbf{400} \\ \bottomrule
\end{tabularx}
\end{table}

\subsection{Evaluation Protocol}
To evaluate the quality of the generated formalizations, we adopt a rigorous two-stage protocol. First, we apply a \textbf{Compiler Verification} filter; any code that fails to compile is immediately assigned a score of 0. Second, for compiling code, we employ an \textbf{LLM-as-a-Judge} (GPT-5.2 Medium Thinking mode) to assess semantic equivalence. 
The evaluator compares the generated Lean statement against the original natural language input and assigns a faithfulness score on a scale of 0--10. 
Because all reported results depend on this judge, we explicitly validate its reliability via cross-judge agreement with an independent model in
(Section~\ref{sec:robustness}).

The judge prompt and rubric (Appendix~\ref{app:judge_rubric}) were reviewed and adjusted by a Lean 4–experienced researcher to align the evaluation with statement-level meaning.

\vspace{0.2cm}
\noindent \textbf{Success Metric:} We define a generated translation $y$ for an input statement $x$ as \textbf{Faithful} if and only if:
\[ \text{Compiles}(y) = \text{True} \quad \textbf{AND} \quad \text{EquivalentScore}(x, y) \ge 9 \]

\section{Methodology}
\label{sec:method}

\begin{figure}[t]
    \centering
    \includegraphics[width=\linewidth]{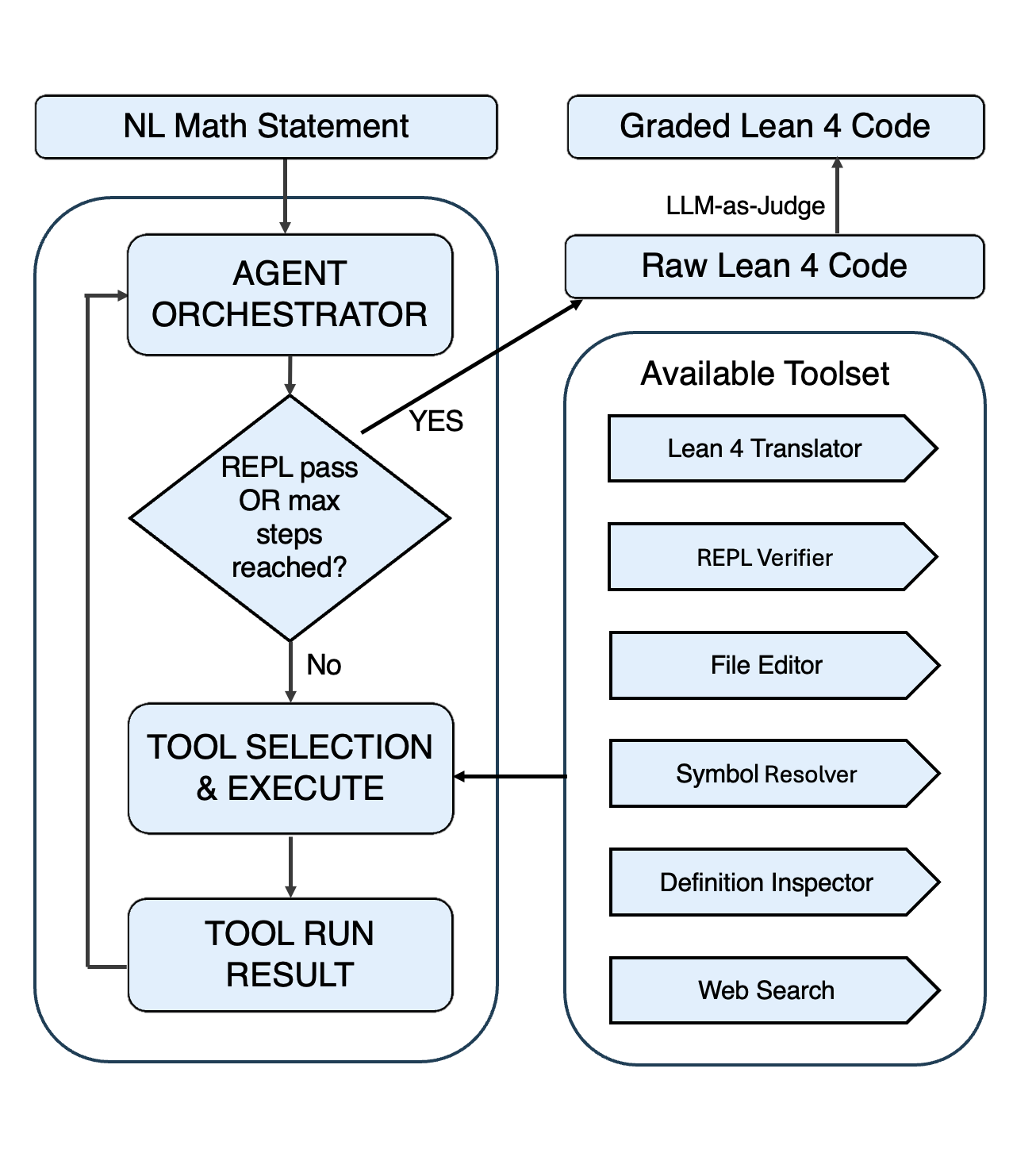} 
    \caption{\textbf{Agent Orchestration Logic.} The architecture consists of a central LLM orchestrator (left) and a Lean 4 execution environment (right).}
    \label{fig:agent_logic}
\end{figure}

The agent architecture, illustrated in Figure~\ref{fig:agent_logic}, is structured as a for-loop. The core component is a central LLM orchestrator (GPT 5.2) that interacts with the Lean 4 environment via a defined API. Unlike static prompting, this architecture allows the model to maintain a persistent state, observing the consequences of its actions—such as compiler errors or retrieval results—before deciding on the next step. This design effectively decouples the LLM reasoning from the Lean Compiler verification.

\subsection{Tool Definitions}
We equip the agent with tools designed to bridge the gap between informal mathematical reasoning and formal Lean 4 syntax:

\begin{itemize}
    \item \textbf{Mathlib Retrieval:} Retrieval tools such as \texttt{lean\_inspect\_name} allow the agent to ask Lean ``what is this symbol?'', returning its real type and definition directly from Mathlib.

    \item \textbf{Expert Drafting:} The \texttt{lean4\_translator} tool requests a translation from the fine-tuned model Herald, providing a specialized starting point.

    \item \textbf{Compiler Feedback:} The \texttt{lean\_repl\_runner} tool exposes the Lean 4 compiler interface. This provides precise error messages, enabling the agent to iteratively diagnose and correct syntax or validity issues.
\end{itemize}

Detailed specifications of the tool definitions and signatures are provided in Appendix~\ref{app:implementation}.

\subsection{Prompt Composition}
\label{subsec:prompt_composition}
To ensure rigorous evaluation, we employ a modular prompting architecture. The system context is dynamically assembled from three independent components, ensuring that the core task definition remains constant while we vary the available tools:

\begin{itemize}
    \item \textbf{Role Definition:} Instructs the agent to act strictly as a translator, not a prover. The objective is to produce a type-correct theorem statement ending in \texttt{:= by sorry}, ignoring proof generation.
    
    \item \textbf{General Guidelines:} Enforces validity standards, such as requiring \texttt{import Mathlib} and strictly prohibiting the invention of new definitions or axioms to prevent "hallucinated" success.
    
    \item \textbf{Tool Specifications:} Contains the functional definitions for the active toolset. We alter the configuration by simply including or excluding tools from this block, without modifying the instructions above.
\end{itemize}

A common failure mode in agent evaluation is mixing up \emph{tool effects} with \emph{prompt tweaks}. If we rewrote the instructions for each tool configuration, any performance difference could come from better prompt wording rather than the tools themselves.

To rule this out, we keep the role definition and general guidelines exactly the same in every experiment. The only thing that changes across configurations is the tool interface block (which tools are declared and how they are described). This makes the factorial ablation in \cref{sec:factorial} cleaner: differences in performance can be attributed to the presence or absence of components such as compiler feedback or expert drafting, rather than to changes in the core instructions.

The full prompt templates, including the shared base prompt and all configuration-specific tool-availability blocks, are provided in Appendix~\ref{app:implementation}.

\subsection{Execution Loop}
Finally, we formalize the interaction between these components as a discrete control loop, shown in Algorithm~\ref{alg:agent_controller}. 
At each step $t$, the agent receives the current message history (including the modular prompt and any tool outputs) and produces either a tool call or a final response. 
This process repeats until the Lean compiler reports success or the step budget $T_{\max}$ is exhausted.

\begin{algorithm}[h]
   \caption{Agentic Controller for Lean Formalization}
   \label{alg:agent_controller}
\begin{algorithmic}[1]
   \State {\bfseries Input:} Statement $x$, Context $p$, ToolSet $\mathcal{T}$
   \State $H \leftarrow [\text{SystemPrompt}, \text{UserMsg}(x, p,\tau)]$
   \For{$t=1$ {\bfseries to} $T_{max}$}
       \State $Msg \leftarrow \text{LLM}(H)$
       \State $H.\text{append}(Msg)$
       \If{$Msg$ calls \textsc{Tool}($name, args$)}
           \State $Result \leftarrow \text{ExecuteTool}(name, args)$
           \State $H.\text{append}(\text{ToolOutput}, Result)$
       \ElsIf{$Msg$ declares \textsc{Success}}
           \State \Return \textsc{Success} \Comment{Compilation Verified}
       \EndIf
   \EndFor
   \State \Return \textsc{Failure}
\end{algorithmic}
\end{algorithm}

\section{Performance Evaluation}
\label{sec:benchmarking}

To quantify the "agentic premium" of our framework, we benchmark the full agent against two static baselines: \textbf{One-Shot Base} (standard GPT-5.2 and Gemini-2.5-Pro prompting without tools) and a \textbf{One-Shot Fine-Tuned} model (Herald), which represents the limit of parametric knowledge without execution feedback.

\subsection{Performance Analysis}

\begin{figure}[h]
    \centering
    \includegraphics[width=0.95\linewidth]{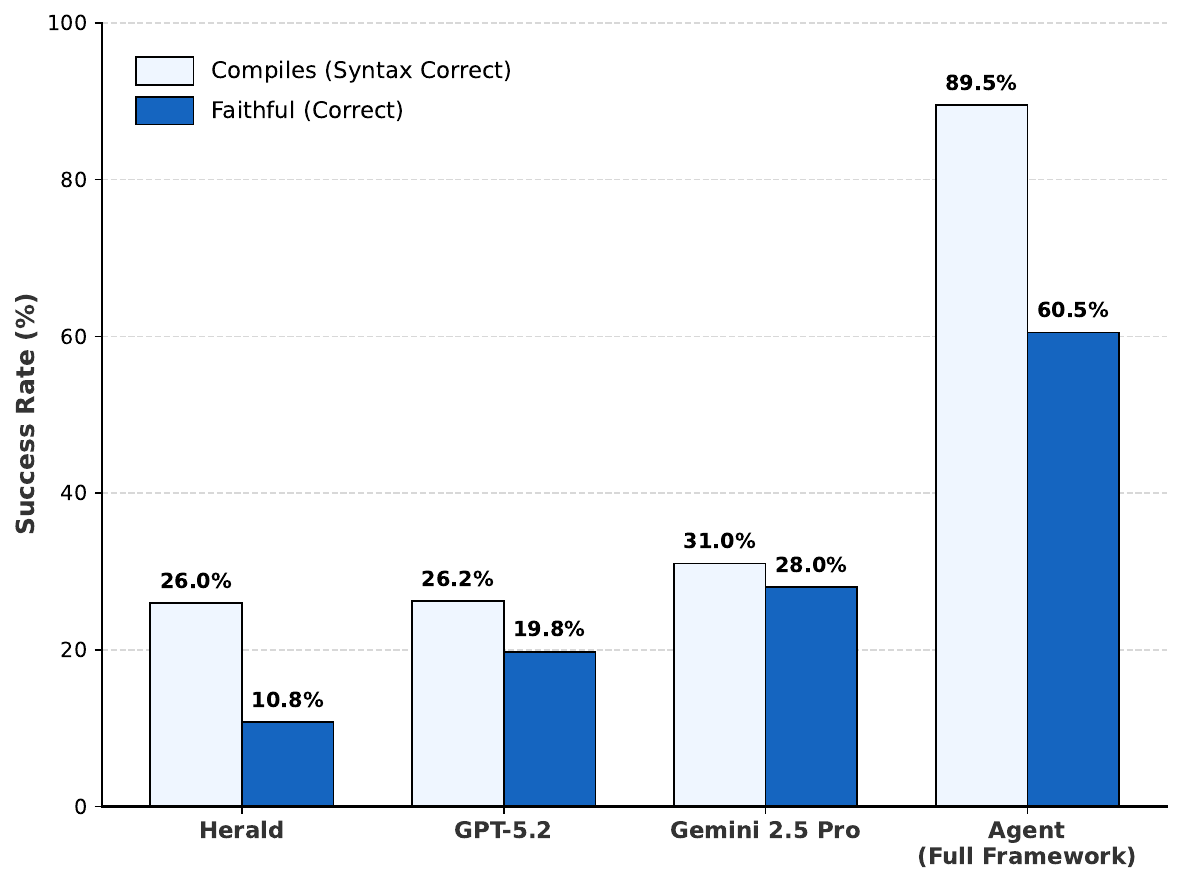} 
    \caption{\textbf{Main Results.} The full framework (all tools enabled; $T_{\max}=24$) achieves substantially higher compilation and faithfulness than one-shot baselines on $N=400$ theorems.}

    \label{fig:main_results}
\end{figure}

Figure~\ref{fig:main_results} presents performance on the 400-theorem benchmark. One-shot generation without tools proves inadequate: Herald and GPT-5.2 achieve compilation rates of only \textbf{26.0\%} and \textbf{26.2\%}, respectively. Faithfulness is even lower, ranging from \textbf{10.8\%} (Herald) to \textbf{28.0\%} (Gemini-2.5-Pro). In contrast, the full tool-augmented agent framework ($T_{\max}=24$) boosts compilation to \textbf{89.5\%} and faithfulness to \textbf{60.5\%}, confirming the value of iterative refinement.

\subsection{Efficiency Analysis}
\label{sec:efficiency}

\begin{figure}[t]
    \centering
    \includegraphics[width=1\linewidth]{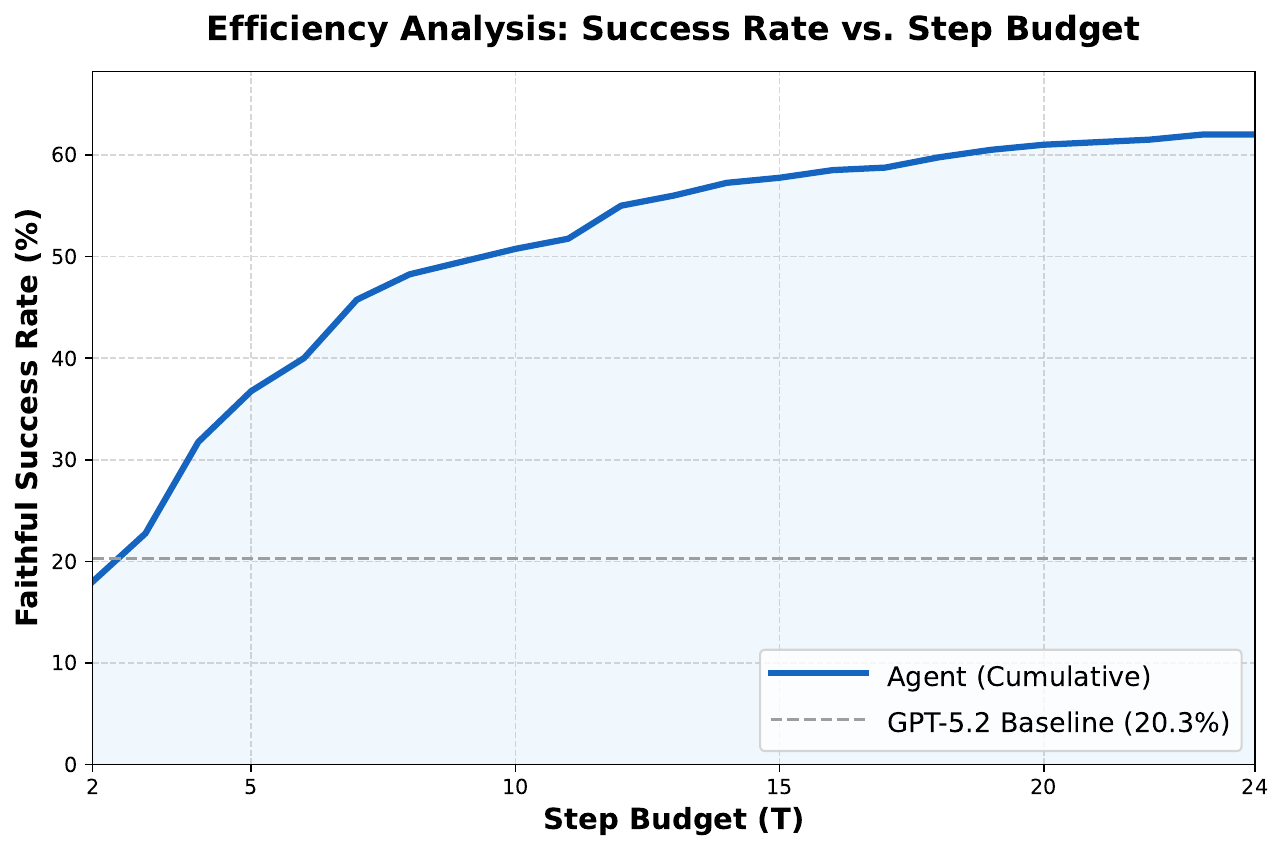} 
    \caption{\textbf{Formalization Efficiency.} Cumulative faithfulness rate as a function of the inference step budget ($T$).}
    \label{fig:efficiency_curve}
\end{figure}

Beyond raw accuracy, and to provide insight into the balance between cost and performance, we analyze the computational budget required to achieve these results. Figure~\ref{fig:efficiency_curve} plots the cumulative success rate against the step budget ($T$), illustrating the trade-off between performance and cost.

The plot reveals two distinct phases in the agent's behavior. We observe rapid convergence up to $T=8$, where nearly half the benchmark is resolved. While returns diminish thereafter, significant gains persist until $T=14$. We therefore identify $T \approx 14$ as the practical saturation point, capturing the bulk of solvable problems before computational costs outweigh improvements.

\subsection{Domain-Specific Analysis}
\label{sec:domain_analysis}

\begin{figure}[t]
  \centering
  \includegraphics[width=\linewidth]{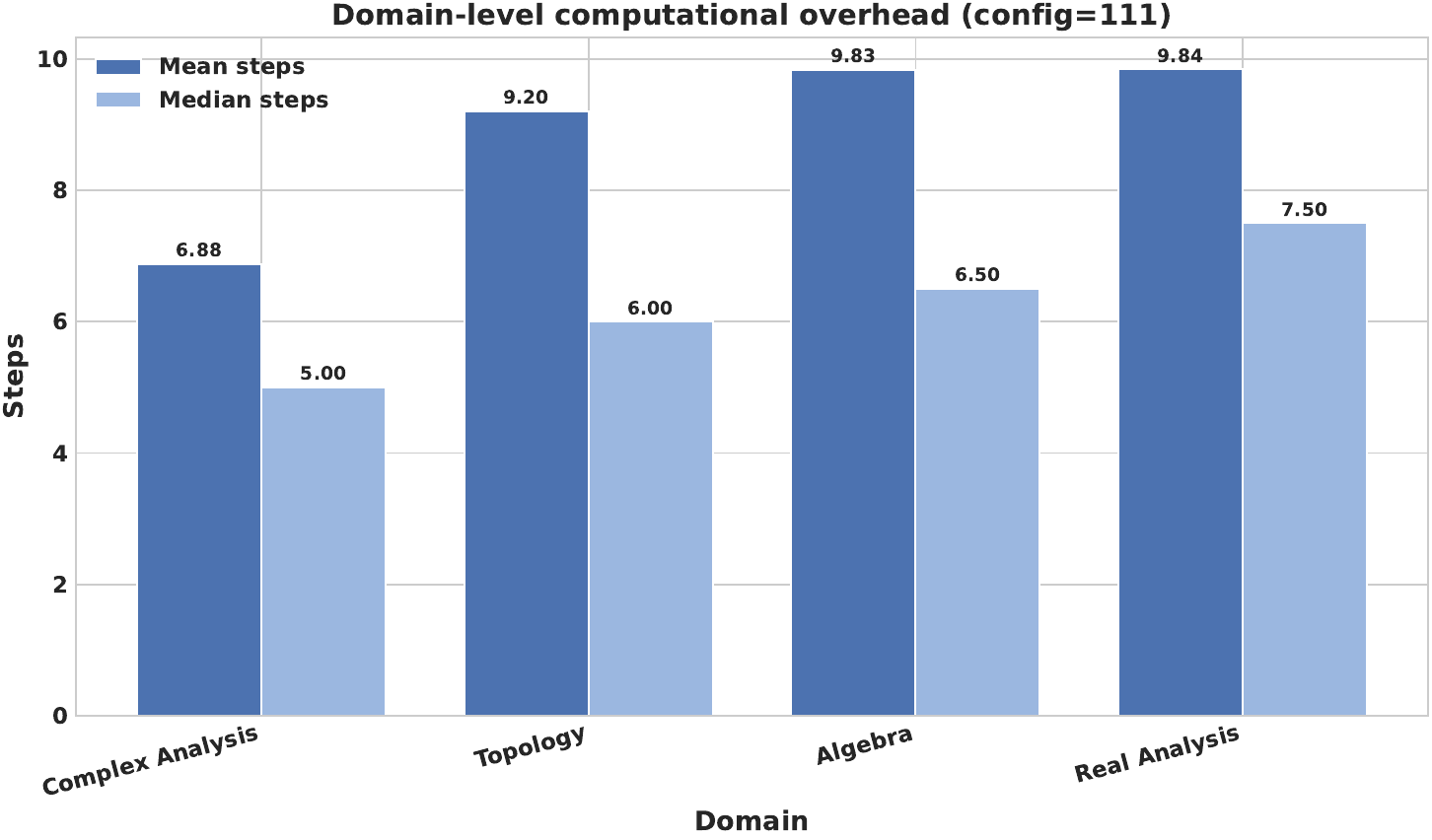}
  \caption{\textbf{Domain-level difficulty (config=111).}
  Mean and median agent steps by domain.}
  \label{fig:domain_analysis}
\end{figure}

Figure~\ref{fig:domain_analysis} reports domain-level computational overhead under full agent framework, measured by mean and median agent steps (N=100 per domain).
\textbf{Complex Analysis} requires the fewest iterations (6.88 mean steps; median 5.0), indicating the lowest computational overhead among the four domains.
\textbf{Real Analysis} is the most expensive (9.84 mean steps; median 7.5), with \textbf{Algebra} close behind (9.83 mean; median 6.5).
\textbf{Topology} exhibits intermediate overhead (9.20 mean; median 6.0).
Appendix~\ref{app:domain_111} and  Appendix~\ref{app:domain_scores} provides the corresponding per-domain summary tables.

\subsection{Model Generalizability}
\label{sec:model_generalizability}

To verify that the framework's gains are not specific to the GPT-5.2 orchestrator, we evaluate two additional orchestrators---Claude Sonnet 4.5 and Gemini-2.5-Pro---on the full benchmark under the 111 configuration. All three converge to 60--65\% consensus faithful despite different one-shot baselines (19--28\%), confirming that the dominance of compiler feedback is structural rather than model-specific. Full results are in Appendix~\ref{app:multi_model}.

\subsection{Robustness and Metric Validation}
\label{sec:robustness}

To validate our LLM-based semantic evaluation, we cross-check the primary judge (GPT-5.2) against another judge (Gemini-2.5-Pro). The two judges exhibit a consistent containment relationship, indicating a structured difference in strictness rather than random disagreement.

\begin{figure}[t] 
\centering \includegraphics[width=0.55\linewidth]{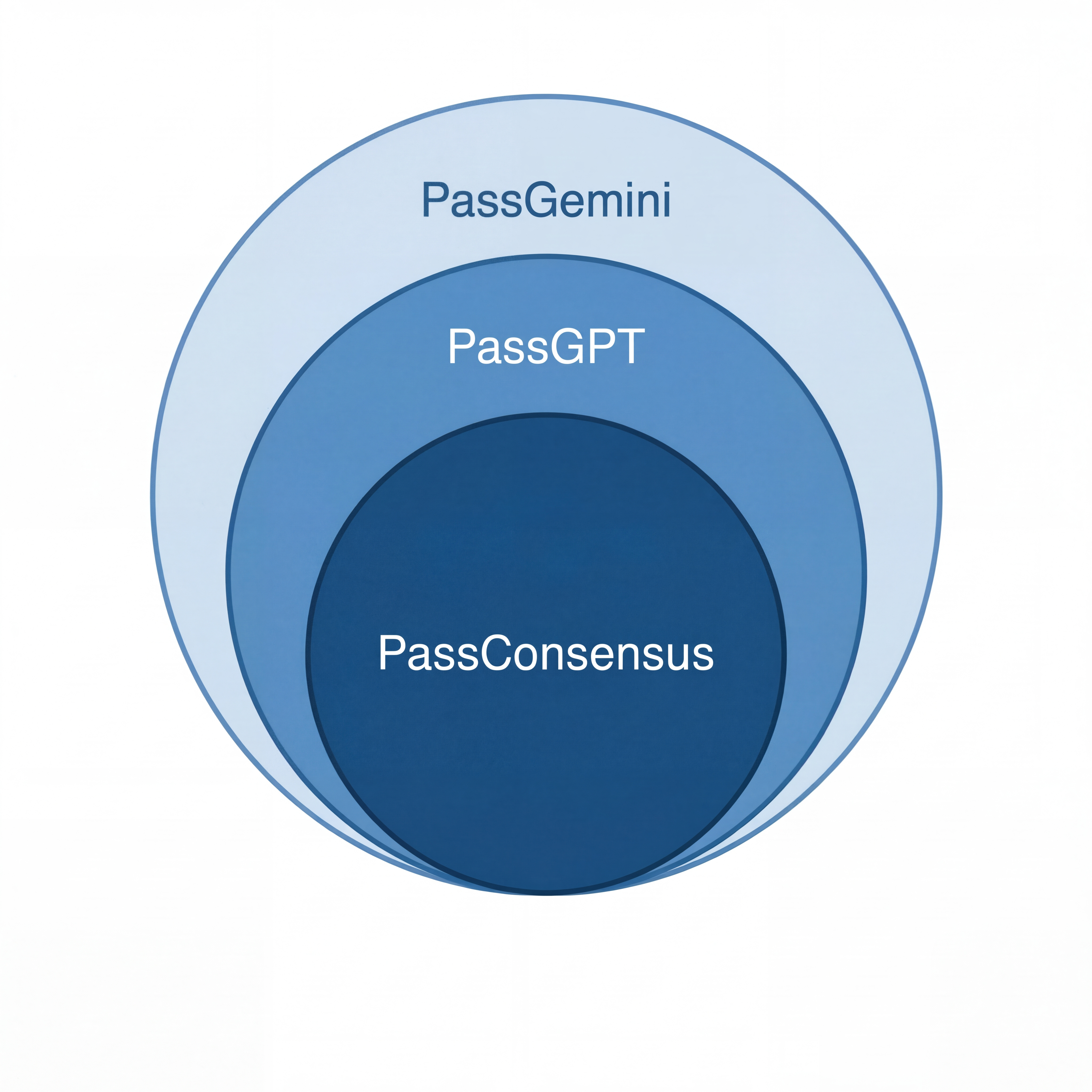} \vspace{-4pt} \caption{\textbf{Inclusion relation for faithfulness labels.}} \vspace{-6pt} \label{fig:consensus_inclusion} \end{figure}

\paragraph{Judge Containment.}
Table~\ref{tab:judge_comparison} reports, for each system, the number of translations judged \emph{Faithful} by each judge. Across all systems, Gemini-2.5-pro consistently accepts more outputs than GPT-5.2, while the consensus is a subset of both. This yields a near-perfect containment pattern
\[
\mathrm{Pass}_{\text{Consensus}} \subseteq \mathrm{Pass}_{\text{GPT}} \subseteq \mathrm{Pass}_{\text{Gemini}}\text{:}
\]
across all ten configurations, at most 6 out of 400 translations are labeled Faithful by GPT-5.2 but not by Gemini, indicating that Gemini-2.5-Pro is systematically more permissive, while GPT-5.2 provides a more conservative semantic filter. The inclusion relation for the consensus-rate definition is shown in Figure~\ref{fig:consensus_inclusion}.

\begin{table}[t]
\centering
\caption{\textbf{Comparison of LLM judges and their consensus.}
Entries count translations that compile and are judged \emph{Faithful} by each judge; \textbf{Consensus} counts those labeled Faithful by both.}
\label{tab:judge_comparison}
\footnotesize
\begin{tabular}{lccc}
\toprule
\textbf{System} & \textbf{Gemini} & \textbf{GPT-5.2} & \textbf{Consensus} \\
\midrule
\multicolumn{4}{l}{\textit{One-shot baselines}} \\
Herald            & 52  & 43  & 43 \\
GPT-5.2           & 90  & 81  & 79 \\
Sonnet 4.5        & 145 & 116 & 114 \\
Gemini-2.5-Pro    & 122 & 112 & 112 \\
\midrule
\multicolumn{4}{l}{\textit{Tool-augmented agents (GPT-5.2 orchestrator)}} \\
Config 100        & 112 & 98  & \textbf{98} \\
Config 001        & 160 & 134 & \textbf{132} \\
Config 101        & 174 & 146 & \textbf{144} \\
Config 110        & 282 & 241 & \textbf{235} \\
Config 111        & 291 & 248 & \textbf{242} \\
Config 010        & 304 & 250 & \textbf{245} \\
Config 011        & 291 & 251 & \textbf{248} \\
\midrule
\multicolumn{4}{l}{\textit{Tool-augmented agents (alt.\ orchestrators, config 111)}} \\
Sonnet 4.5        & 315 & 271 & \textbf{262} \\
Gemini-2.5-Pro    & 307 & 263 & \textbf{262} \\
\bottomrule
\end{tabular}
\end{table}

\paragraph{Conservatism of the Primary Metric.}
We quantify agreement using the \emph{consensus rate}, defined as the fraction of GPT-5.2-labeled Faithful translations that are also labeled Faithful by Gemini:
\[
\mathrm{ConsensusRate}
:=
\frac{\#\mathrm{Pass}_{\text{Consensus}}}{\#\mathrm{Pass}_{\text{GPT}}}.
\]
Across all baselines and tool-augmented agents, the consensus rate exceeds 97\% in every case, with an overall average of \textbf{98.7\%}. For example, on the full system (111), GPT-5.2 labeled 248 translations as Faithful, of which Gemini also labeled 242 as Faithful (97.6\%). Together with the containment pattern above, this supports using the strict GPT$\cap$Gemini consensus as the primary faithfulness metric throughout the paper (Section~\ref{sec:problem}), providing a conservative lower bound on semantic faithfulness.

\begin{figure}[t]
    \centering
    \includegraphics[width=0.9\linewidth]{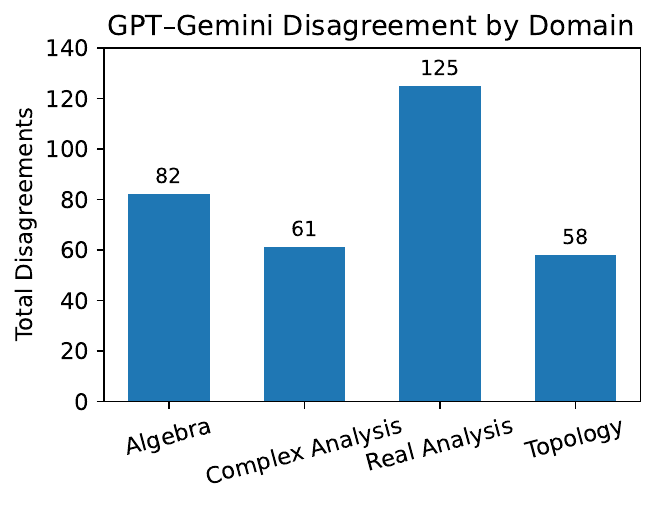}
    \caption{\textbf{GPT--Gemini disagreement by mathematical domain.}
    Total number of GPT--Gemini disagreements aggregated across all ten experimental systems.}
    \label{fig:judge_disagreement_domain}
\end{figure}

\paragraph{Where do judges disagree?}
Figure~\ref{fig:judge_disagreement_domain} breaks down GPT--Gemini disagreements by mathematical domain, aggregated across all ten experimental systems (three one-shot baselines and seven tool-augmented agents). Each domain contains 100 problems evaluated under 10 systems, so each bar summarizes 1{,}000 judged translations.

Although overall agreement is high (Section~\ref{sec:robustness}), the remaining disagreements are not uniformly distributed across domains. Real Analysis accounts for the largest number of conflicts (125), followed by Algebra (82), Complex Analysis (61), and Topology (58). This structured concentration suggests that judge disagreements are systematically associated with particular mathematical domains. This pattern aligns with our domain difficulty results: under the full agent setting (config=111), Real Analysis exhibits the lowest Faithful rate among the four domains (Appendix~\ref{app:domain_extra}, Table~\ref{tab:domain_111_metrics}), suggesting that harder domains leave more borderline cases for judges to interpret.

\paragraph{Human Expert Validation.}
To directly validate the LLM-as-Judge metric, a Lean~4 expert independently reviewed 138 translations that the consensus metric classified as \emph{Faithful}, scoring each on the same 0--10 rubric.
Of 138 scored translations, 83.3\% (115/138) received exactly matching scores between the human and LLM judges.
A further 14.5\% (20/138) disagreed by $\pm1$ within the faithful range (scores of 9 vs.\ 10), leaving only 3 out of 138 (2.2\%) where the human score crossed below the faithful threshold ($\geq$9)---all cases where the LLM overestimated (LLM scored 10, human scored 5--7).
This yields 97.8\% binary agreement on the faithful/unfaithful classification, confirming that the automated metric is well-calibrated with a slight positive bias. 

We note two limitations of this audit: (1)~it covers only outputs classified as faithful (score $\geq$9), so it measures precision but not recall; and (2)~the expert was not blinded to the LLM scores. However, the expert did not uniformly confirm the LLM labels---in 3 of 138 cases, substantially lower scores were assigned---suggesting independent judgment rather than mechanical agreement. A broader audit covering the full score distribution is left to future work.

\section{Factorial Analysis}
\label{sec:factorial}

While the superior performance of the agent is evident, the ``black box'' nature of the full system obscures which components drive this success. To disentangle these factors, we conduct a systematic factorial analysis to quantify \emph{why} the agent succeeds.

\subsection{Experimental Design: Factorial Ablation}
\label{subsec:exp_design}

To systematically isolate the contribution of each component, we decompose the agent arsenal into three tool factors:
\begin{itemize}
  \item \textbf{T (Translation expert).} Access to the fine-tuned Herald (7B) model for translation.
  \item \textbf{F (Feedback).} Access to the Lean compiler feedback loop (REPL) for error correction and compile status.
  \item \textbf{S (Search).} Symbol-retrieval tools (retrieval/web search) for resolving unknown definitions.
\end{itemize}

We adopt a full $2^3$ factorial design to evaluate interaction effects. We explicitly map the empty-set configuration ($\emptyset$) to the One-Shot Baseline established in Section~\ref{sec:benchmarking}, as an agentic loop with zero tools conceptually reduces to standard autoregressive generation. The remaining seven configurations are evaluated by selectively toggling tool definitions in the system prompt.

\subsection{Results and Interpretation}
\label{subsec:factorial_results}

Table~\ref{tab:factorial_results} reports performance across all configurations, grouped by the presence of Compiler Feedback (\textbf{F}).
To quantify the contribution of each factor, we compute standard factorial main and interaction effects, defined as differences in mean response between the high and low levels of a factor (averaged over the other factors).

\begin{table}[t]
\centering
\caption{\textbf{Factorial analysis results.} Performance across all $2^3$ configurations ($N=400$), grouped by the presence of the REPL feedback tool (\textbf{F}). We use 0 to denote that tool is turned off.}
\label{tab:factorial_results}
{\footnotesize
\resizebox{\columnwidth}{!}{%
\begin{tabular}{ccc|cc|c}
\toprule
\multicolumn{3}{c|}{\textbf{Config}} & \multicolumn{2}{c|}{\textbf{Performance}} & \textbf{Gain} \\
\textbf{T} & \textbf{F} & \textbf{S} & \textbf{Comp. (\%)} & \textbf{Faith. (\%)} & $\Delta$Faith. (pts) vs .\ 000 \\
\midrule
\addlinespace
\multicolumn{6}{l}{\textit{Regime 1: $F=0$ (Agent with no REPL)}} \\
\addlinespace
0 & 0 & 0 & 26.25 & 19.75 & -- \\
1 & 0 & 0 & 30.25 & 24.50 &  +4.75\\
0 & 0 & 1 & 45.50 & 33.00 &  +13.25\\
1 & 0 & 1 & 50.00 & 36.00 &   +16.25\\

\midrule
\addlinespace
\multicolumn{6}{l}{\textit{Regime 2: $F=1$ (Agent with REPL)}} \\
\addlinespace
1 & 1 & 0 & 93.50 & 58.75 &  +39.00\\
1 & 1 & 1 & 89.50 & 60.50 & +40.75\\
0 & 1 & 0 & 91.50 & 61.25 &  +41.50\\
0 & 1 & 1 & 87.25 & 62.00 &  +42.25\\

\bottomrule
\end{tabular}
}}
\end{table}

\paragraph{Main-Effect Definition.}
Let $Y(T,F,S)$ denote the Faithful accuracy for configuration $(T,F,S)$ in Table~\ref{tab:factorial_results}.
For any factor $X \in \{T,F,S\}$, its factorial main effect is
\begin{equation}
\label{eq:main_effect}
\text{Effect}(X)
=
\mathbb{E}[\,Y \mid X{=}1\,]
-
\mathbb{E}[\,Y \mid X{=}0\,],
\end{equation}
where the expectation averages uniformly over all settings of the other two factors.
Table~\ref{tab:main_effects} reports the resulting main effects.

\begin{table}[t]
\centering
\caption{\textbf{Factorial main effects on Faithful accuracy.}
Each effect is computed by Eq.~\ref{eq:main_effect}. 95\% confidence intervals are obtained via bootstrap resampling ($B=10{,}000$).}
\label{tab:main_effects}
{\footnotesize
\setlength{\tabcolsep}{8pt}
\renewcommand{\arraystretch}{1.05}
\begin{tabular}{lcccc}
\toprule
\textbf{Factor} & $X{=}1$ & $X{=}0$ & \textbf{Effect} & \textbf{95\% CI} \\
\midrule
Feedback (F)    & 60.6 & 28.3 & \textbf{+32.3} & [28.7, 35.9] \\
Search (S)      & 47.9 & 41.1 & \textbf{+6.8}  & [3.6, 10.0] \\
Translation (T) & 44.9 & 44.0 & +0.9            & [$-$2.1, 4.0] \\
\bottomrule
\end{tabular}
}
\end{table}

\begin{itemize}
    \item \textbf{Feedback is the capability bottleneck ($+32.3$ pts):} Enabling the REPL transforms the agent from a low-success regime (19.75--36.00\%) into a high-success regime (58.75--62.00\%). This shows that iterative compiler-guided repair is the dominant mechanism enabling faithful Lean formalization. Notably, the magnitude of this effect varies by domain, with Complex Analysis benefiting most (+53 pts) and Algebra/Topology least (+20 pts); see Appendix~\ref{app:domain_feedback} for details.

    \item \textbf{Search is a stabilizer ($+6.8$ pts):} Symbol-level retrieval via \texttt{\#check}/\texttt{\#print} provides a consistent boost by reducing hallucinations and improving name resolution, even though it does not create a new performance regime on its own. Domain-level analysis (Appendix~\ref{app:domain_search}) shows that Search yields +12.4 pts when full REPL is absent, but near-zero gain (+1.2 pts) when Feedback is enabled---consistent with the negative $F{\times}S$ interaction (Section~\ref{subsec:interaction_effects}), as full compiler diagnostics subsume symbol-level queries.
    
    \item \textbf{Drafting is largely substitutable ($+0.9$ pts):} The specialized Herald model yields only a small average improvement. Because the orchestrator (GPT-5.2) is already strong, external drafting becomes redundant once feedback and search are available, and in some cases slightly interferes with the repair loop.
\end{itemize}

\subsection{Interaction Effects: Capability vs.\ Efficiency}
\label{subsec:interaction_effects}

Main effects summarize average marginal gains, but they can hide strong \emph{regime dependence}.
To expose when tools improve \emph{final faithfulness} versus primarily \emph{efficiency}, we compute
\emph{simple effects} conditioned on the presence of compiler feedback $F$.

\paragraph{When does Search improve final faithfulness?}
Define the Search simple effect at fixed feedback level $F{=}f$ as
\[
\Delta_S(F{=}f)
=
\mathbb{E}[Y \mid S{=}1, F{=}f]
-
\mathbb{E}[Y \mid S{=}0, F{=}f],
\]
where the expectation averages uniformly over $T$.
From Table~\ref{tab:factorial_results}, Search yields a large gain without REPL feedback,
$\Delta_S(F{=}0)=+12.4$ points, but only a marginal gain with REPL feedback,
$\Delta_S(F{=}1)=+1.2$ points.
Thus, Search meaningfully improves faithfulness in the low-capability regime ($F=0$), but in the high-capability regime ($F=1$) it mainly shifts \emph{how} the agent acquires information---from repeated compile--repair iterations toward earlier, targeted symbol queries---consistent with the reduced REPL-call counts in Table~\ref{tab:tool_usage}.

\paragraph{When does drafting hurt?}
Similarly, define the Translator simple effect
\[
\Delta_T(F{=}f)
=
\mathbb{E}[Y \mid T{=}1, F{=}f]
-
\mathbb{E}[Y \mid T{=}0, F{=}f],
\]
averaging over $S$.
Drafting helps when feedback is absent ($\Delta_T(F{=}0)=+3.9$ points) but hurts when feedback is present ($\Delta_T(F{=}1)=-2.0$ points).
This ``drafting penalty'' suggests that, once compiler-guided repair is available, specialist drafts may become redundant or introduce anchoring effects that impede convergence to the most faithful formulation.

\paragraph{Substitution captured by interactions.}
The regime dependence above appears as negative interaction terms under the standard difference-in-differences definition (averaging over the remaining factor): $F{\times}S=-11.1$ points and $F{\times}T=-5.9$ points, while $S{\times}T\approx 0$.
In our implementation, both $F$ (whole-program compilation) and $S$ (targeted symbol queries) are backed by the same installed Lean/Mathlib snapshot; in particular, $S$ is realized via Lean-level checks such as \texttt{\#check}/\texttt{\#print} rather than an independent external knowledge source.
Under this setup, the negative interactions are consistent with overlapping functionality between whole-program diagnostics and symbol-level lookup: when REPL feedback is enabled, enabling Search is associated with only small changes in final faithfulness, while shifting some information gathering from compiler-error-driven trial-and-error toward earlier identifier and type resolution. A domain-level breakdown of this effect is provided in Appendix~\ref{app:domain_search}.

\paragraph{Scale and Transfer Effects.}
A plausible explanation for the limited marginal value of the Expert Drafter ($T$) is the disparity in model capacity and pretraining diversity. While the Herald translator is fine-tuned on Lean data, the orchestrator (GPT-5.2) is trained at much larger scale on a broad mixture of programming languages, mathematics, and natural language. This large-scale, multi-domain pretraining enables strong cross-domain transfer, which—when combined with compiler feedback and retrieval—allows the generalist model to adapt to Lean syntax and semantics through iterative correction, reducing the marginal value of a domain-specialized draft.

\paragraph{The value of verification beyond retrieval.}
Search provides symbol-level semantic primitives (existence checks, types, and definitions via \texttt{\#check}/\texttt{\#print}), but it does not provide whole-statement verification or full elaboration diagnostics. In contrast, Feedback adds (i) a global compilation/elaboration signal and (ii) actionable error messages that enable iterative repair, which appear to be the dominant drivers of faithful formalization.

\subsection{Behavioral Analysis: REPL Efficiency}
\label{subsec:repl_efficiency}

While the factorial analysis in Section~\ref{subsec:factorial_results} identifies which tools determine final Faithful accuracy, tool invocation logs provide complementary insight into \emph{how efficiently} the agent reaches that accuracy. In particular, the number of calls to the Lean REPL (\texttt{lean4\_repl\_runner}) serves as a proxy for the amount of trial-and-error required to obtain compilable and faithful Lean code.

\begin{table}[t]
\centering
\caption{\textbf{Tool Usage Counts (Total).}
Cumulative tool invocations across the subset with complete tool-call logs ($N=384$; all accuracy results use $N=400$).
We report REPL calls (\texttt{lean4\_repl\_runner}) and aggregate symbol-retrieval tools into
$S_{\text{total}}=\texttt{search\_online}+\texttt{lean\_inspect\_name}+\texttt{lean\_resolve\_name}$.}

\label{tab:tool_usage}
\setlength{\tabcolsep}{6pt}
\renewcommand{\arraystretch}{1.1}
{\footnotesize
\begin{tabular}{c|rrr|l}
\toprule
\textbf{$(T,F,S)$} & \textbf{Trans.} & \textbf{REPL} & \textbf{$S_{\text{total}}$} & \textbf{Notes} \\
\midrule
(0,1,0) & 0   & 1496 & 0    & Feedback-only \\
(0,1,1) & 0   & 1050 & 1726 & +Search (REPL$\downarrow$) \\
\midrule
(1,1,0) & 257 & 1374 & 0    & Feedback + Expert \\
(1,1,1) & 112 & 1008 & 1913 & +Search (REPL$\downarrow$) \\
\bottomrule
\end{tabular}
}
\end{table}

As shown in Table~\ref{tab:tool_usage}, using the convention $(T,F,S)$, adding Search reduces REPL usage in both $T=0$ and $T=1$ settings when feedback is enabled. For $T=0$, REPL calls drop from 1{,}496 in (010) to 1{,}050 in (011), a 29.8\% reduction. For $T=1$, REPL calls drop from 1{,}374 in (110) to 1{,}008 in (111), a 26.6\% reduction. Meanwhile, Search usage increases substantially (from 0 to 1{,}726 or 1{,}913), indicating that the agent shifts effort from expensive compile-and-repair iterations toward cheaper symbol-level validation.

One plausible mechanism is that, before Search is available, the agent may rely on REPL feedback not only for “compile/no-compile” verification but also largely to infer missing identifiers, expected variables, and type constraints from Lean’s error messages; introducing Search could reduce this reliance by providing symbol-level information earlier.

Finally, we note that the extensive interaction logs generated by our agent constitute a valuable dataset of successful and failed formalization trajectories. These traces map the trial-and-error process of resolving compiler errors, making them an ideal corpus for future research into Reinforcement Learning (RL) for formalization. By training on these repair loops, future models might internalize the feedback dynamics, learning to predict successful repairs without incurring the high inference cost of the live REPL.

\section{Limitations}
\label{sec:limitations}

Our study has several important limitations that contextualize the interpretation of the results.

\paragraph{Statement Fidelity vs.\ Provability.}
Our task is translating natural language into valid Lean~4 statement declarations (ending in \texttt{:= by sorry}). We enforce syntactic validity and semantic faithfulness but do not verify \emph{provability}. In a small pilot on 12 benchmark problems, Aristotle achieved 11/12 faithful translations by consensus, while our agent achieved 8/12 on the same subset using general-purpose LLMs. A systematic comparison with Lean-specialized models remains future work.

\paragraph{Dependence on the Lean Environment.}
In our current implementation, all formal feedback ultimately flows through the Lean 4 environment, either directly via REPL compilation or indirectly through symbol-level queries backed by the installed Mathlib snapshot. As a result, our factorial analysis primarily reflects differences in \emph{interaction modality} (whole-program compilation vs.\ targeted symbol lookup) rather than access to fundamentally independent sources of mathematical truth.

A natural extension is to build a static, agent-facing Mathlib dataset (e.g., indexed statements, docstrings, and symbol metadata) that supports offline lookup and reduces reliance on the live REPL during search. However, this direction introduces its own limitation: Mathlib evolves rapidly, so any fixed dataset can become stale, potentially yielding outdated names, moved theorems, or changed type signatures unless it is versioned and regularly refreshed.

\paragraph{Restricted Mathematical Coverage.}
Our benchmark covers four major domains (Real Analysis, Complex Analysis, Topology, Algebra). We caution that these results may not generalize to combinatorial or discrete domains (e.g., Graph Theory, Number Theory), where formalization relies less on structural type definitions and more on intricate logical predicates.

\paragraph{Compute Cost.}
Agentic formalization is significantly more expensive than one-shot generation. Although symbol search tools reduce the number of compiled iterations, the inference cost of the full REPL loop remains a barrier to real-time interactive assistance.

\section{Conclusion}
\label{sec:conclusion}

We presented a systematic study of tool-augmented agents for translating natural-language mathematics into Lean~4. By combining a generalist LLM with retrieval, expert drafting, and compiler feedback, the agent raises faithfulness from 28.0\% to 60.5\% over the strongest one-shot baseline on a new 400-theorem benchmark.

Beyond raw performance, our factorial analysis reveals a clear structural picture. Compiler feedback is the dominant capability driver, enabling a transition from unreliable drafts to consistently faithful formalizations. Symbol-level search improves efficiency and stability by reducing hallucinations and lowering the number of expensive compile--repair cycles, while expert drafting provides only marginal gains once feedback and search are available. These results demonstrate that execution and retrieval, rather than parametric specialization alone, are the critical ingredients for scalable mathematical formalization.

Together, these findings suggest a general design principle for formal reasoning agents: high-precision execution environments should serve as the primary scaffold for LLM reasoning, with retrieval used to stabilize and accelerate convergence. Our results suggest a broader lesson: for rapidly evolving formal libraries, one-off fine-tuning on a static corpus may be brittle, and sustained progress likely depends more on verification-coupled iteration (and continual refresh) than on a fixed offline drafter alone. We believe this paradigm will be central to future systems that aim to bridge informal mathematical knowledge and machine-verifiable proof.

\section*{Impact Statement}
This work advances methods for translating informal mathematics into formal proof assistant statements. Potential impacts include improving reliability and accessibility of formal verification tools. Risks include overreliance on automated formalization and evaluation bias; we advocate transparent reporting of tool configurations and failure modes.

\bibliography{references}
\bibliographystyle{icml2026}

\newpage
\appendix
\onecolumn

\section{Agent Implementation Details}
\label{app:implementation}

\subsection{Prompt Assembly}
The system prompt is constructed from three modular components:
(i) a \textbf{task definition} establishing the core translation objective,
(ii) a \textbf{general usage guide} containing fixed best practices for Lean~4 code generation, and
(iii) a \textbf{capability block} that is dynamically populated based on the factorial configuration $(T,F,S)$.
This structure holds the agent’s semantic goal and coding standards (i and ii) fixed, while selectively enabling or disabling external tools (iii) to isolate the causal effect of each capability.
We assemble (i) and (ii) into a \textbf{Shared Base Prompt}, shown below, while the capability block definitions are provided in Appendix~\ref{app:tool_blocks}.

\subsection{Shared Base Prompt}
\label{app:base_prompt}

\begin{tcolorbox}[
  colback=white,
  colframe=gray,
  title=Base System Prompt,
  listing only,
  breakable,
  listing options={
    basicstyle=\footnotesize\ttfamily,
    breaklines=true,
    columns=fullflexible
  }
]
You are an expert Lean4 translation agent.

Your task is to translate a natural-language mathematical statement into
faithful Lean4 syntax (NOT a proof).
The final result must:
- import Mathlib at the very top
- compile in Mathlib
- be semantically faithful to the original statement
- end with `:= by sorry`

You are NOT proving anything.
You are only producing a correctly typed, correct-meaning Lean statement.

GENERAL INSTRUCTIONS FOR CODE GENERATION

• The Lean file MUST start with:
  import Mathlib

• The Lean file MUST contain exactly ONE final translated statement
  representing the original natural-language meaning.

• The final statement MUST end with:
  := by sorry

• Do NOT write any proof code before `:= by sorry`
  (no `by`, `simp`, `have`, `calc`, etc. anywhere before the final `:= by sorry`).

• Do NOT invent new definitions, axioms, constants, or placeholder structures.

• Prefer robust formulations:
  use quantifiers, membership, and $\leftrightarrow$ characterizations rather than fragile definitional equalities.

• Only finish when the last written version:
  (1) compiles in Mathlib (if lean4\_repl\_runner is available)
  (2) is semantically faithful to the original statement

When both are satisfied, return:
{ "status": "success" }
\end{tcolorbox}

\subsection{Tool-Availability Blocks}
\label{app:tool_blocks}

Each factorial configuration $(T,F,S)$ determines which tools are exposed to the agent. We implement this by a modular prompt template in which the \texttt{AVAILABLE TOOLS} block is constructed by including the corresponding tool-specification entries and omitting inactive ones. We show the complete tool block for the fully enabled configuration $(1,1,1)$ below; other configurations are obtained by deleting the entries for tools that are disabled.

\begin{tcolorbox}[
  colback=white,
  colframe=gray,
  title={Tool Block (T=1, F=1, S=1)},
  listing only,
  breakable,
  listing options={
    basicstyle=\footnotesize\ttfamily,
    breaklines=true,
    columns=fullflexible
  }
]
AVAILABLE TOOLS

\smallskip
\texttt{lean4\_translator(statement)} \\
\hspace*{1em} Draft a Lean 4 statement using the fine-tuned Herald translator (you may edit or ignore).

\texttt{lean\_write\_file(code)} \\
\hspace*{1em} Write the full Lean file to the workspace (imports + exactly one final statement).

\texttt{lean4\_repl\_runner()} \\
\hspace*{1em} Compile the current Lean file and return compiler feedback.

\texttt{lean\_inspect\_name(name, imports?, include\_print?)} \\
\hspace*{1em} Query Mathlib about an identifier via \texttt{\#check}/\texttt{\#print}.

\texttt{lean\_resolve\_name(token, namespace\_hints?, imports?, top\_k?)} \\
\hspace*{1em} Suggest likely Mathlib identifiers for an unknown or ambiguous token.

\texttt{search\_online(query)} \\
\hspace*{1em} Run a web search and return results for the query.

\end{tcolorbox}

\subsection{Tool Interfaces}
\label{app:tool_interfaces}

\begin{tcolorbox}[
  colback=white,
  colframe=black!50,
  title=Agent Tool API,
  breakable
]
\centering
\footnotesize
\setlength{\tabcolsep}{5pt}
\renewcommand{\arraystretch}{1.25}
\begin{tabular}{p{3.5cm} p{4.5cm} p{6.5cm}}
\toprule
\textbf{Tool} & \textbf{Arguments} & \textbf{Functionality} \\
\midrule

\texttt{lean\_inspect\_name} 
& \texttt{name}, \texttt{imports?}, \texttt{include\_print?} 
& Checks whether a Mathlib symbol exists and returns its type
(\texttt{\#check}) and optionally its definition (\texttt{\#print}). \\

\texttt{lean\_resolve\_name}
& \texttt{token}, \texttt{namespace\_hints?}, \texttt{imports?}, \texttt{top\_k?}
& Performs fuzzy symbol resolution to map informal or incorrect names
to valid Mathlib identifiers. \\

\texttt{lean4\_repl\_runner}
& \texttt{Lean 4 code to be executed}
& Compiles the current Lean file and returns compiler diagnostics
(errors, warnings, and success flags). \\

\texttt{lean4\_translator}
& \texttt{statement}
& Invokes the fine-tuned Herald translator to generate a draft Lean
formalization of the natural-language input. \\

\texttt{search\_online}
& \texttt{query}
& Performs external web search to retrieve mathematical or
documentation context not present in Mathlib. \\

\bottomrule
\end{tabular}
\end{tcolorbox}

\subsection{Example Execution Trace}
\label{app:trace}

\begin{tcolorbox}[
  colback=white,
  colframe=black!50,
  title=\textbf{Example Execution Trace (agent tool calls)},
  breakable
]
\footnotesize
\setlength{\tabcolsep}{6pt}
\renewcommand{\arraystretch}{1.25}
\begin{tabular}{@{}p{0.9cm}p{3.2cm}p{11.2cm}@{}}
\toprule
\textbf{Step} & \textbf{Tool} & \textbf{Arguments (agent side)} \\
\midrule

1 & \texttt{lean\_write\_file} &
\texttt{path="jirilebl\_ca\_ca\_17658.lean"}; \;
\texttt{content = import Mathlib ... theorem jirilebl\_ca\_ca\_17658 ... := by sorry} \\

2 & \texttt{lean4\_repl\_runner} &
\texttt{path="jirilebl\_ca\_ca\_17658.lean"} \\

3 & \texttt{lean\_inspect\_name} &
\texttt{name="Polynomial.IsConstant"} \\

4 & \texttt{lean\_inspect\_name} &
\texttt{name="Polynomial.eval"} \\

5 & \texttt{lean\_inspect\_name} &
\texttt{name="Polynomial.isConstant"} \\

6 & \texttt{lean\_inspect\_name} &
\texttt{name="Polynomial.natDegree"} \\

7 & \texttt{lean\_write\_file} &
\texttt{path="jirilebl\_ca\_ca\_17658.lean"}; \;
\texttt{content = patched theorem uses p.natDegree $\neq$ 0 and Polynomial.eval} \\

8 & \texttt{lean4\_repl\_runner} &
\texttt{path="jirilebl\_ca\_ca\_17658.lean"} \\

\bottomrule
\end{tabular}
\end{tcolorbox}

\section{LLM-as-a-Judge Prompt and Rubric}
\label{app:judge_rubric}

We evaluate semantic faithfulness using an LLM-as-a-judge that receives:
(i) the natural-language statement, (ii) the generated Lean 4 code, and (iii) a boolean \texttt{compile\_pass}.
The judge outputs a JSON object with fields \texttt{faithful} and \texttt{grade} (0--10).
If \texttt{compile\_pass=false}, we require \texttt{faithful=false} and restrict the grade to 0--3. The translation is considered as faithful if and only if the code compiles and the score is $>=9$.

\begin{tcolorbox}[
  title=Judge system prompt (verbatim),
  colback=white,
  colframe=gray,
  breakable
]
\begin{Verbatim}[breaklines=true,breakanywhere=true]
You are an expert in Lean 4, Mathlib, and mathematics. You are judging TRANSLATION-ONLY.

Input: (1) a natural-language statement, (2) a Lean 4 code snippet, (3) compile_pass boolean.
Your job: decide whether the Lean code, AS A STATEMENT, matches the meaning of the natural-language statement.

Key policy (NOT PICKY):
- If compile_pass = False: the translation is NOT faithful. grade must be 0..3. faithful=false.
- If compile_pass = True: ignore the proof/body entirely (including `by sorry`). Proof completeness is NOT part of the evaluation.
- A translation is faithful if the Lean statement expresses the same mathematical claim as the NL statement.

Auxiliary definitions policy (lenient but not allowing cheating):
- Auxiliary defs/lemmas are allowed if they are reasonable encodings/abbreviations and do not change the meaning.
- However, if the code introduces a clearly vacuous placeholder for a nontrivial concept (e.g. `def X := True`, `:= none`, `:= 0` for something meant to be meaningful), and that placeholder is essential to making the final theorem appear to match, then the translation is NOT faithful.

How to judge meaning (focus):
- Compare the MAIN theorem/definition statement(s) to the NL statement.
- Check quantifiers (forall/exists), logical structure (->/<->/and/or), and key hypotheses.
- Check main objects/domains: Nat/Int/Real, rings/groups, ZMod n, matrices, etc.
- Small implementation details are OK if the meaning is preserved.

Scoring guide (integer 0..10):
- 0: unrelated.
- 1-3: compile_pass is False OR statement is clearly wrong.
- 4-6: compiles, but meaning is materially different / missing key hypotheses / wrong domain;
       might be "in the ballpark".

- 7-8: compiles, mostly matches, but has a noticeable mismatch (e.g. strengthened/weakened in an important way).
- 9: compiles, very close; only tiny mismatch.
- 10: compiles and meaning matches.

Output contract (STRICT):
Return a single JSON object with exactly these fields:
{
  "faithful": true or false,
  "grade": 0..10,
  "thought": "### BEGIN THOUGHT\n<short explanation focusing on statement-level comparison>\n### END THOUGHT"
}
Return ONLY valid JSON. No extra keys. No markdown outside JSON.
\end{Verbatim}
\end{tcolorbox}

\section{Additional Domain Results}
\label{app:domain_extra}

\subsection{Domain Metrics under Full Configuration}
\label{app:domain_111}

\begin{tcolorbox}[title=Domain metrics under config 111, colback=white, colframe=gray, breakable]
\small
\begin{table}[H]
\centering
\caption{\textbf{Domain metrics under config=111.}
Conditional faithfulness is computed as Faithful/Compile.}
\label{tab:domain_111_metrics}
\setlength{\tabcolsep}{4.5pt}
\begin{tabular}{lrrrrr}
\toprule
\textbf{Domain} & \textbf{Compile} & \textbf{Faithful} & \textbf{Faithful|Compile} & \textbf{Steps (mean)} & \textbf{Steps (median)} \\
\midrule
Complex Analysis & 0.95 & 0.82 & 0.86 & 6.88 & 5 \\
Real Analysis    & 0.89 & 0.49 & 0.55 & 9.84 & 7.5 \\
Algebra          & 0.87 & 0.56 & 0.64 & 9.83 & 6.5 \\
Topology         & 0.87 & 0.61 & 0.70 & 9.20 & 6 \\
\bottomrule
\end{tabular}
\end{table}
\end{tcolorbox}

\subsection{Judge Scores by Domain}
\label{app:domain_scores}

\begin{tcolorbox}[title=Average judge scores by domain across configurations, colback=white, colframe=gray, breakable]
\small
\begin{table}[H]
\centering
\caption{\textbf{Average faithfulness scores by domain and configuration.}
Columns correspond to tool configurations in $(T,F,S)$ bit order.}
\label{tab:domain_score_all}
\setlength{\tabcolsep}{4.5pt}
\begin{tabular}{lrrrrrrr}
\toprule
\textbf{Domain} & \textbf{001} & \textbf{010} & \textbf{011} & \textbf{100} & \textbf{101} & \textbf{110} & \textbf{111} \\
\midrule
Algebra          & 5.60 & 7.14 & 7.23 & 4.46 & 5.85 & 7.20 & 7.22 \\
Complex Analysis & 4.72 & 9.27 & 8.91 & 4.20 & 4.94 & 8.94 & 8.98 \\
Real Analysis    & 4.62 & 7.38 & 7.34 & 3.59 & 4.75 & 7.33 & 7.31 \\
Topology         & 5.39 & 7.10 & 7.03 & 4.47 & 6.14 & 7.05 & 7.25 \\
\bottomrule
\end{tabular}
\end{table}

\noindent\textbf{Interpretation.}
Score trends are consistent with the domain difficulty ranking in Section~\ref{sec:domain_analysis}:
Complex Analysis attains the highest scores under tool-enabled settings, while Real Analysis remains lowest.
\end{tcolorbox}

\subsection{Domain-Level Feedback Effects}
\label{app:domain_feedback}

\begin{figure}[H]
\centering
\includegraphics[width=0.7\textwidth]{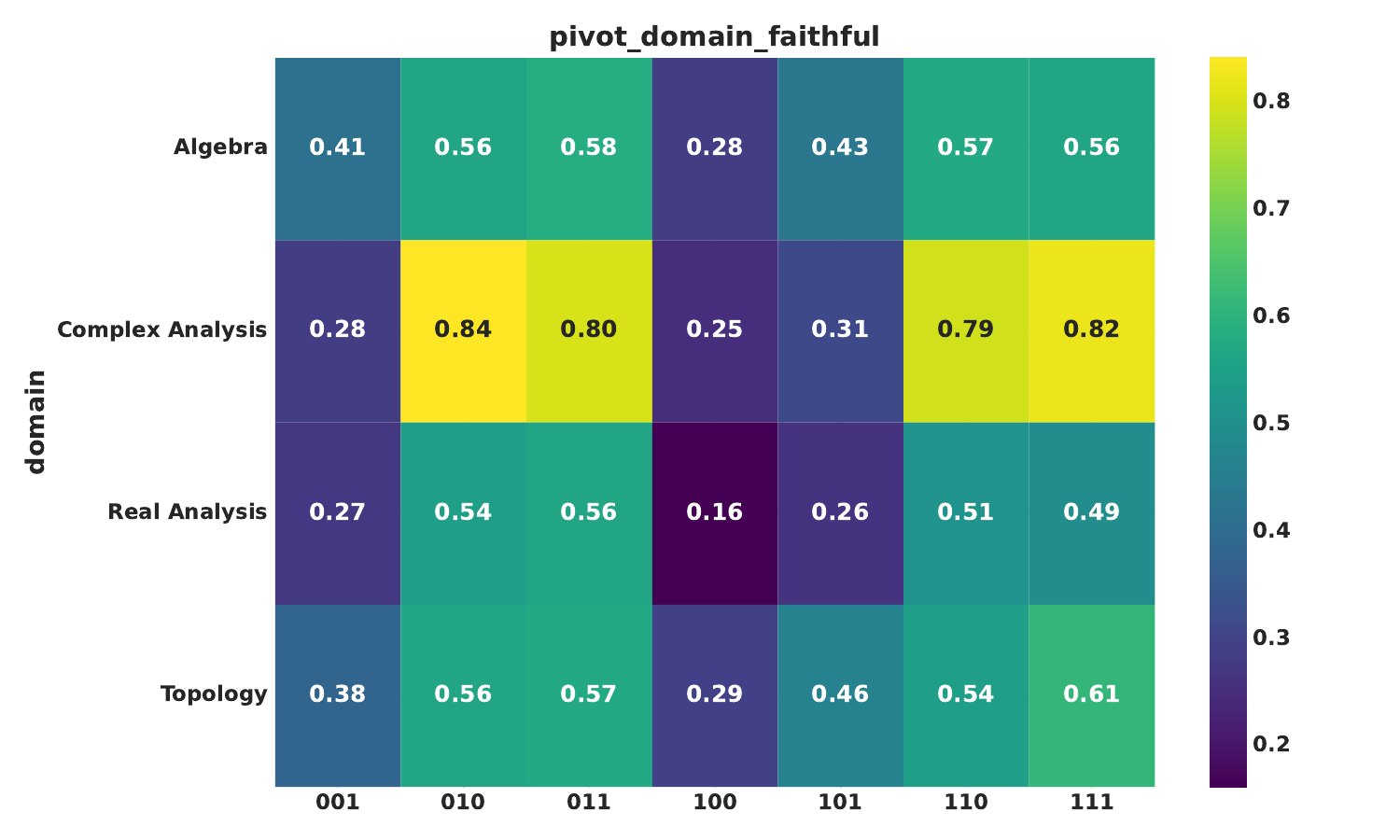}
\caption{\textbf{Faithful rate by domain and configuration.}}
\label{fig:domain_faithful_heatmap}
\end{figure}

\begin{tcolorbox}[title=Domain-wise effect of Compiler Feedback (F), colback=white, colframe=gray, breakable]
\small
\begin{table}[H]
\centering
\setlength{\tabcolsep}{8pt}
\begin{tabular}{lrrr}
\toprule
\textbf{Domain} & \textbf{F=0 (avg)} & \textbf{F=1 (avg)} & \textbf{$\Delta$F} \\
\midrule
Complex Analysis & 0.28 & 0.81 & \textbf{+0.53} \\
Real Analysis    & 0.23 & 0.53 & +0.30 \\
Algebra          & 0.37 & 0.57 & +0.20 \\
Topology         & 0.38 & 0.57 & +0.19 \\
\bottomrule
\end{tabular}
\end{table}

\noindent\textbf{Interpretation.} Complex Analysis benefits most from compiler feedback (+53 pts), likely due to mature Mathlib coverage. Algebra and Topology show smaller gains (~+20 pts), suggesting either limited library support or intrinsic formalization difficulty.
\end{tcolorbox}

\subsection{Domain-Level Effects of Search (S)}
\label{app:domain_search}

\begin{tcolorbox}[title=Simple effect of Search conditioned on Feedback, colback=white, colframe=gray, breakable]
\small

The Search tool (S) provides symbol-level queries via \texttt{\#check}/\texttt{\#print}, accessing a subset of REPL functionality without full compilation diagnostics. As noted in Section~\ref{subsec:interaction_effects}, both tools are backed by the same Lean/Mathlib snapshot, leading to the negative interaction $F{\times}S=-11.6$ pts.

\begin{table}[H]
\centering
\setlength{\tabcolsep}{6pt}
\begin{tabular}{lrr|rr}
\toprule
 & \multicolumn{2}{c|}{\textbf{F=0}} & \multicolumn{2}{c}{\textbf{F=1}} \\
\textbf{Domain} & \textbf{S=0} & \textbf{S=1} & \textbf{S=0} & \textbf{S=1} \\
\midrule
Algebra          & 0.28 & 0.42 & 0.565 & 0.57 \\
Complex Analysis & 0.25 & 0.30 & 0.815 & 0.81 \\
Real Analysis    & 0.16 & 0.27 & 0.525 & 0.525 \\
Topology         & 0.29 & 0.42 & 0.55  & 0.59 \\
\midrule
\textbf{Avg $\Delta$S} & \multicolumn{2}{c|}{\textbf{+10.8 pts}} & \multicolumn{2}{c}{\textbf{+1.0 pts}} \\
\bottomrule
\end{tabular}
\end{table}

\noindent\textbf{Interpretation.} When full REPL is unavailable (F=0), symbol queries provide substantial gains (+10.8 pts average). Once full REPL is enabled (F=1), the marginal benefit of Search drops to near zero (+1.0 pts), confirming that whole-program diagnostics subsume symbol-level information. This domain-level breakdown corroborates the negative $F{\times}S$ interaction reported in Section~\ref{subsec:interaction_effects}.
\end{tcolorbox}

\subsection{Multi-Model Orchestrator Comparison}
\label{app:multi_model}

To assess whether the framework's gains are specific to the GPT-5.2 orchestrator or reflect structural properties of tool-augmented formalization, we evaluate two additional orchestrator models---Claude Sonnet 4.5 and Gemini-2.5-Pro---on the full 400-theorem benchmark under the 111 (all tools) configuration.

\begin{figure}[H]
    \centering
    \includegraphics[width=0.55\linewidth]{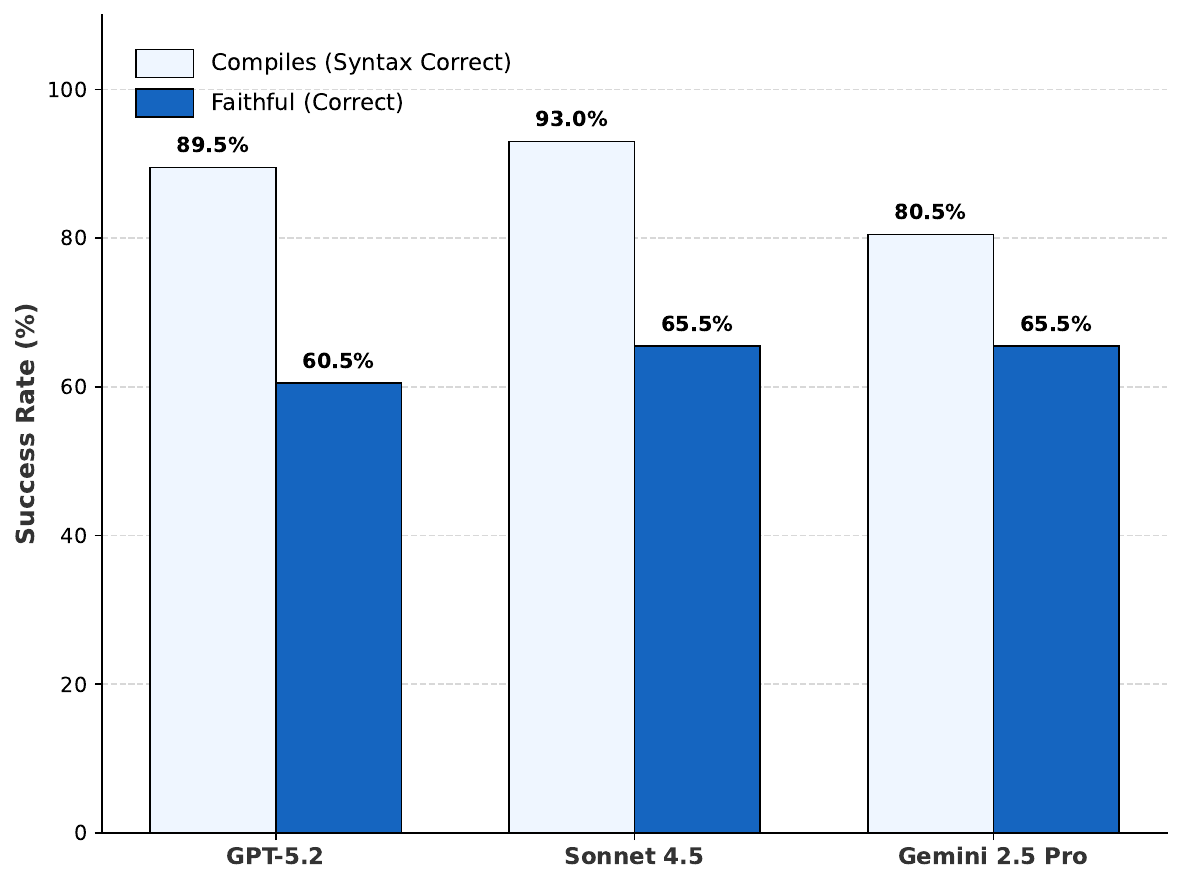}
    \caption{\textbf{Multi-model agent comparison (config 111).} All three orchestrators converge to 60--65\% consensus faithful despite different one-shot baselines (19--28\%).}
    \label{fig:multi_model}
\end{figure}

As shown in Figure~\ref{fig:multi_model}, all three models show large uplift from the tool-augmented framework (+148 to +163 faithful translations) and converge to 60--65\% consensus faithful despite markedly different one-shot baselines (19--28\%). This convergence confirms that the findings of the factorial analysis---particularly the dominance of compiler feedback---are structural rather than model-specific.

\end{document}